\documentclass[twocolumn,prl,showpacs,floatfix,superscriptaddress]{revtex4}
\usepackage{graphicx}
\usepackage{amsfonts}
\usepackage{amssymb}
\usepackage{bm}

\newcommand{\be}{\begin{equation}}
\newcommand{\ee}{\end{equation}}
\newcommand{\bea}{\begin{eqnarray}}
\newcommand{\eea}{\end{eqnarray}}
\newcommand{\zncu}{ZnCu$_3$(OH)$_6$Cl$_2$}

\begin{document}

\title{Magnetic Susceptibility of the Kagome Antiferromagnet \zncu}

\author{Marcos Rigol}
\affiliation{Department of Physics and Astronomy,
University of Southern California, Los Angeles, California 90089, USA}
\author{Rajiv R.~P.~Singh}
\affiliation{Department of Physics, University of California, Davis,
CA 95616, USA}

\date{\today}


\begin{abstract}
We analyze the experimental data for the magnetic susceptibility of the 
material \zncu\ in terms of the Kagome Lattice Heisenberg model (KLHM), 
discussing possible role of impurity spins, dilution, exchange anisotropy, 
and both out-of-plane and in-plane Dzyaloshinsky-Moriya (DM) anisotropies, 
with explicit theoretical calculations using the Numerical Linked Cluster 
(NLC) method and exact diagonalization (ED). The high-temperature experimental 
data are well described by the pure Heisenberg model with $J=170 K$. We show 
that the sudden upturn in the susceptibility around $T=75 K$ is due to DM 
interactions. We also observe that at intermediate temperatures, below $T=J$, 
our calculated susceptibility for KLHM fits well with a power law $T^{-0.25}$. 
\end{abstract}

\maketitle

Frustrated magnetic systems represent outstanding challenges in condensed 
matter physics \cite{expt-review}. While much progress has been made theoretically 
over the last few decades \cite{theory}, key issues related to concrete lattice 
models remain unresolved. Computational methods are restricted to small system 
sizes or high temperatures, whereas the starting point for field theory methods, 
which maybe asymptotically exact at low temperatures, involve drastic approximations, 
which makes their validity for realistic lattice models unclear. Experiments can 
bridge this gap by providing information all the way from high to low temperatures.

The spin-half Kagome-Lattice Heisenberg Model (KLHM), is one of the best 
studied frustrated quantum spin models. Much of our knowledge of the system comes 
from studies of finite-size periodic clusters. Exact Diagonalization studies 
\cite{lhuillier} suggest that the model has a short spin-spin correlation length, 
and a spin gap of approximately $J/20$. In the finite systems, there are a large 
number of singlet states below the lowest triplet and their number grows with the 
system size \cite{mila}. Various Resonating Valence Bond (RVB) scenarios have been 
invoked, and the finite size spectra have also been interpreted to imply deconfined 
spin-half excitations.

In light of these theoretical results, the experimental behavior of the recently 
synthesized Kagome-Lattice materials ZnCu$_3$(OH)$_6$Cl$_2$ are highly unexpected
\cite{expts1,expts2}. The high temperature inverse susceptibility data was found 
to obey a Curie-Weiss law, with a Curie-Weiss constant of about $300K$. Yet, there 
appears no spin-gap either in the susceptibility or the specific heat or the neutron 
spectra or in the nuclear spin-lattice relaxation $T_1$, down to temperatures below 
$100 mK$. At the lowest temperatures the susceptibility saturates and the specific 
heat shows power-law behavior in temperature. The latter is suppressed by magnetic 
fields, showing it to be magnetic in origin. A possible interpretation of the 
substantial rise in susceptibility at low temperatures is that it is due to impurity 
spins outside the Kagome planes, caused by substitutions of non-magnetic $Zn$ sites 
with $Cu$ [7]. However, the fact that the muon shift $K$ tracks the bulk susceptibility 
$\chi$, implies that the susceptibility is not due to inhomogeneities localized 
away from the Kagome planes, but rather a bulk behavior of the system.

Here, we use the triangle-based NLC method \cite{nlc} and ED, to calculate the 
magnetic susceptibility of the KLHM including various perturbations such as 
dilution, exchange anisotropy and Dzyaloshinsky-Moriya anisotropy. We find that 
the high temperature susceptibility is indeed in excellent agreement with that 
of KLHM. However, the Curie-Weiss fit is fortuitous. The true Curie-Weiss behavior 
(with $T_{CW}=J$) is only valid for $T>10J$, a region inaccessible to experiments. 
When Curie-Weiss fits are made to data for $T<2J$, the effective Curie-Weiss constant 
increases in units of $J$, and hence the true $J$ is smaller than reported in Ref.
\cite{expts1,expts2}. The high temperature data fits well with $J=170K$.

The sharp upturn in the susceptibility around $75 K$ cannot be accounted for by 
the KLHM. We find that only a Dzyaloshinsky-Moriya (DM) anisotropy can account for 
the sharp upturn in the experimental susceptibility. A way to distinguish this from 
an impurity spin would be by the anisotropy. The excess susceptibility due to DM 
would be strongly anisotropic, and this anisotropy is temperature dependent.

The NLC method allows us to calculate the susceptibility for the KLHM accurately 
down to $T=0.3J$, without extrapolations. In contrast, the high temperature 
expansions (HTE) fail to converge below $T=J$ \cite{elstner}. Some extrapolations 
of NLC and HTE suggest that shortly below $T=0.3J$, the susceptibility begins to 
turn down \cite{elstner,nlc}. Unfortunately, these extrapolations to lower 
temperatures become subjective, and hence we focus only on the region $T>0.3J$. 
All NLC calculations reported below are done using the triangle based expansion, 
where contributions to the thermodynamic system from graphs with complete triangles 
are included \cite{nlc}. As appropriate, two highest order calculations are shown.

The Heisenberg Hamiltonian in field $h$ is
\begin{equation}
{\cal H}=J\sum_{\langle i,j\rangle} {\bf S}_i {\bf S}_j - g\mu_B h\sum_i S^z_i.
\end{equation}
In our calculations we set $J=1$, Boltzmann constant $k=1$ and $g\mu_B=1$. 
The susceptibility per spin is 
\begin{equation}
\chi=\frac{T}{N}\left.\frac{\partial^2 \ln{Z}}{\partial h^2}\right\vert_{h=0},
\end{equation}
where $Z$ is the partition function. The molar susceptibility, measured experimentally, 
is related to our susceptibility per spin by the relation, $\chi_{molar}= C \chi$,
where the constant $C=N_A g^2\mu_B^2/kJ=0.3752 g^2/J$ in cgs units. At high 
temperatures the susceptibility has a Curie-Weiss form with Curie-Weiss temperature 
$T_{\rm cw}=J/k$. As discussed by Zheng {\it et~al.} \cite{zheng}, asymptotic 
Curie-Weiss behavior is only valid for $T\gg J$. At temperatures of order $J$, 
one can define an effective temperature dependent Curie-Weiss constant 
$T_{\rm cw}^{\rm eff}$ as
\begin{equation}
T_{\rm cw}^{\rm eff} =-T - {\chi\over d\chi/dT}
\end{equation}
If one was to fit $\chi^{-1}$ to a linear curve in the vicinity of some temperature 
$T$ and use the intercept to estimate the Curie-Weiss constant, one would get 
$T_{\rm cw}^{\rm eff}$. This quantity for KLHM is shown in Fig.~1. Note that for 
$T<2J$ one has $T_{\rm cw}^{\rm eff}\approx 2J$. This means that $J$ for 
\zncu\ is smaller than estimated in Refs.\ \cite{expts1,expts2}. In the inset of 
Fig.~1, we show a comparison of the experimental susceptibility with the theoretical 
one with $J=170K$. The agreement is excellent at high temperatures.

\begin{figure}[!htb]
\begin{center}
  \includegraphics[scale=.55,angle=0]{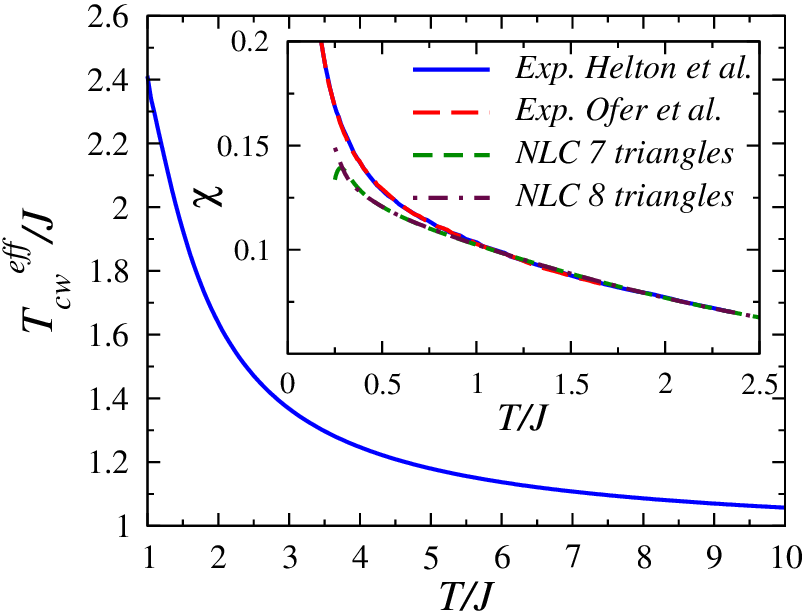}
\end{center}
\vspace{-0.6cm}
\caption{\label{EffectiveCurieWeiss} Temperature dependent effective Curie-Weiss
temperature $T_{cw}^{eff}$ for the KLHM. The inset shows fits to the experimental
data of Ref.\ \cite{expts1,expts2}. For both experimental cases we have 
taken $J=170 K$. The $g$-factors needed for the fit are $2.19$ for the data 
from Ref.\ \cite{expts1} and $2.33$ for the data from Ref.\ \cite{expts2}. 
On the theoretical side, the two
highest order NLC results are shown.}
\end{figure}

While the theoretical susceptibility for the KLHM has a weak upturn 
at $T$ near $0.3J$, it is inconsistent with the pronounced rise in 
the measured susceptibility, which grows significantly below $75 K$ 
and only saturates to a much higher value at temperatures below $1K$. 
The experimental behavior could arise from several 
sources. First, there maybe external impurity spins, such as those caused by 
substitution of $Zn$ sites with $Cu$ \cite{ran-06}, which are only very weakly 
coupled to the electronic spins in the Kagome planes. If present, they would add 
a Curie like $1/T$ term to the susceptibility at high temperatures. Fig.~2 shows 
that considering $3$ percent impurities allows us to extend the fit down to $T=0.3J$,
but there remains a sharp rise at lower temperatures.
A $6$ percent impurity concentration eliminates the upturn but makes 
the data inconsistent with the KLHM. Hence, free impurity spins by themselves, 
cannot account for the rise in the data. Besides, as argued by Ofer {\it et~al.}
\cite{expts1}, in this scenario, it would be difficult to understand why the 
muon-shift tracks the bulk susceptibility. One would expect local probes to 
show inhomogeneous behavior and deviate from the bulk averages.

\begin{figure}[!htb]
\begin{center}
  \includegraphics[scale=.55,angle=0]{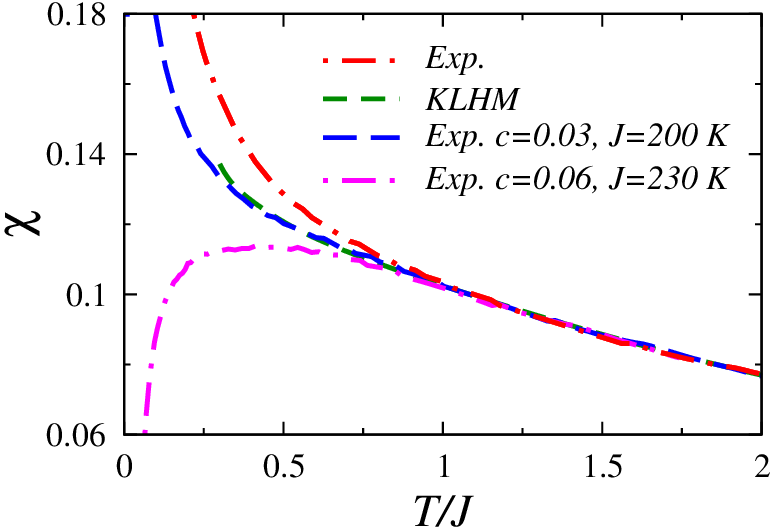}
\end{center}
\vspace{-0.6cm}
\caption{\label{SpinImpurities}
Comparison of KLHM susceptibility with experimental susceptibility after 
subtracting Curie contributions from $3$ and $6$ percent impurities.}
\end{figure}

A second potential reason for the sudden ferromagnetic behavior could be dilution 
due to substitution of $Cu$ sites in the Kagome-planes by $Zn$. The missing spins 
on the lattice could create local moments in the singlet background and cause a 
Curie-like susceptibility to arise at low temperatures. They could weakly couple 
to each other and hence saturate at some much lower temperature. In order to 
investigate this possibility, we have studied the KLHM with quenched dilution. 
We assume that at each site, we have a hole with probability $c$ and a spin with
probability $1-c$. The holes are fixed in their position and extensive quantities 
are averaged over the random configurations $C$, using the relation
\begin{equation}
\langle O\rangle=\sum_{C} P(C) O(C)
\end{equation}
where $P(C)$ is the probability of the configuration $C$. The susceptibility for 
several values of hole concentration is shown in Fig.~3. We note that at these 
intermediate temperatures, holes simply lower the susceptibility and do not lead 
to any ferromagnetic tendencies. Thus the static holes do not provide an explanation 
for the experimental data, either.
We have also calculated the susceptibility of KLHM with 
exchange anisotropy of Ising and XY type. Neither can account for the 
experimental behavior.

\begin{figure}[!htb]
\begin{center}
  \includegraphics[scale=.55,angle=0]{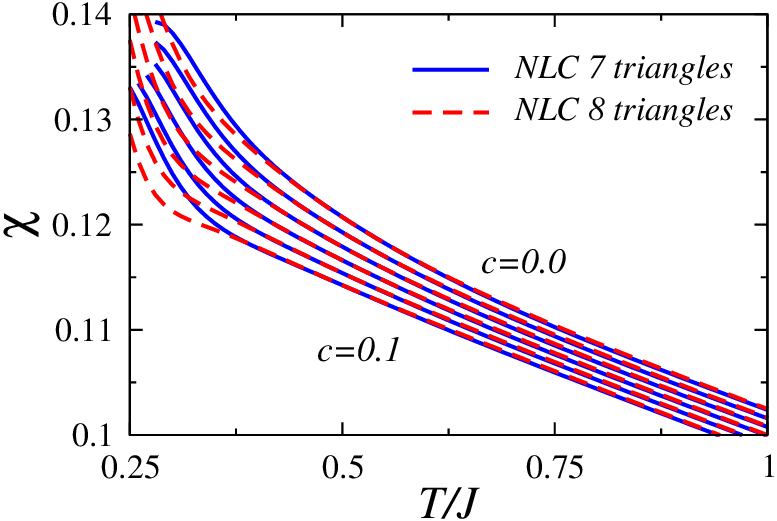}
\end{center}
\vspace{-0.6cm}
\caption{\label{QHSusceptibility.eps}
NLC results for the magnetic susceptibility of the KLHM with hole 
concentration $c$ ranging from $0$ to $10$ percent.}
\end{figure}

We now turn to the Dzyaloshinsky-Moriya (DM) interaction. In Kagome magnets
(which we assume lies in the $x-y$ plane), both out of plane ($D_z$) and 
in plane ($D_p$) DM terms are allowed \cite{harris,yamabe}.
\begin{equation}
{\cal H}_{DM}=\sum_{\langle i,j\rangle}  D_z ({\bf S_i}\times{\bf S_j})_z
         +{\bf D_p}\cdot ({\bf S_i}\times{\bf S_j}),
\end{equation}
The $D_z$ term alternates between the up and down pointing triangles of the Kagome 
lattice. Its sign can be set by demanding that for the up pointing triangle
shown in Fig.~4b, a positive $D_z$ multiplies $({\bf S_1}\times{\bf S_2})_z$.
The in-plane DM term $D_p$ is perpendicular to the bonds and 
points inward towards the center of the triangles \cite{harris,yamabe}. 
Since the DM 
terms break spin rotational symmetry, we need to calculate separately the 
susceptibility with field along $z$ ($\chi_z$) and in the $x-y$ plane ($\chi_p$). 
The powder susceptibility $\chi_a$ 
is given by $\chi_a={1\over 3}(2\chi_p+\chi_z)$. 

With the DM terms, the different $S^z$ sectors are coupled so we are able
to do NLC calculations only up to 6 triangles \cite{nlc}. Unfortunately,
the convergence is poor, and hence we turn to ED of clusters with 12 and 15 
sites (with periodic boundary conditions) \cite{elstner} 
to study the effects of DM 
interactions. We find that a pure $D_z$ term supresses the in-plane and $z$ 
susceptibilities with respect to KLHM. On the other hand, a pure $D_p$ enhances 
both in-plane and z susceptibilities with respect to KLHM. Hence, the competition 
between $D_z$ and $D_p$ can produce varying results.
Once $D_p\neq 0$ the sign of $D_z$ is also relevant 
\cite{elhajal02}. In all cases studied, the DM terms produce $z$ 
susceptibilities which are larger than the in-plane ones. 

As long as $D_p>|D_z|$, we find an enhancement of the powder susceptibility 
with respect to the KLHM. As an example, we show in Figs.~4(a) and 4(b) 
results for the powder susceptibilities and anisotropy when $D_p=0.3J$ 
and $D_z=-0.15J,-0.3J$, respectively. Note that in the theoretical 
calculations the $g$-factor is assumed to be isotropic. Since the 
$z$-susceptibility rises very rapidly, an anisotropic $g$-factor enhanced 
along $z$ will cause an even more rapid rise and will lead to agreement 
with experiments with a smaller DM anisotropy \cite{anisotropy}.

Our main point with Fig.\ 4 is to show that even though more 
information from the material \zncu\ may be required in order to make 
a precise estimate of the DM anisotropy, DM interactions are essential
to understanding that material. We predict that single crystal 
measurements should see an upturn in the anisotropy when the powder 
susceptibility departs from the KLHM result. Such a behavior has already 
been seen for spin-$5/2$ Kagome system KFe$_3$(OH)$_6$(SO$_4$)$_2$ \cite{grohol}.

\begin{figure}[!htb]
\begin{center}
  \includegraphics[scale=.55,angle=0]{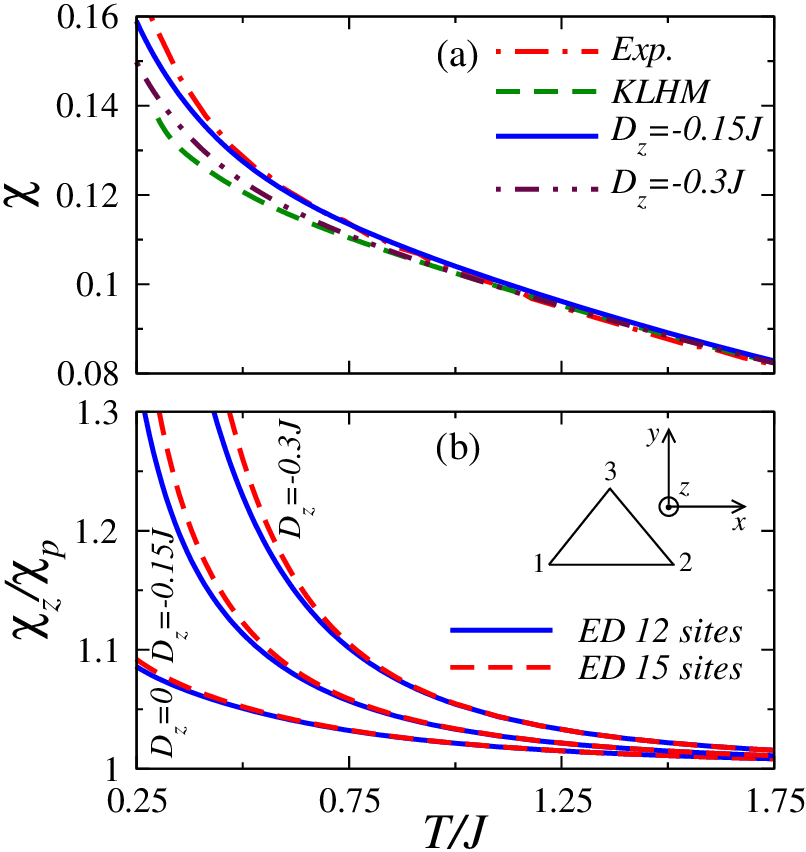}
\end{center}
\vspace{-0.5cm}
\caption{\label{DMSusceptibility.eps}
(a) 15 sites ED calculation of the powder susceptibility with a DM 
term $D_p=0.3J$ and different values of $D_z$, compared with experiments 
and pure KLHM. (b) Anisotropy vs temperature
for $D_p=0.3J$ and different values of $D_z$.
}
\end{figure}

Here, we cannot address the measurements at $T\ll J$. An
interesting question to ask is whether the lack of spin-gap seen in
the experiments is due to DM anisotropy. In 
magnetically ordered systems, the DM term often creates a spin-gap in the 
spin-wave spectra. The KLHM is an interesting system where the spin-gap maybe 
non-zero but singlet excitations maybe gapless \cite{lhuillier,mila}. In that 
case, the DM term can cause the spin-gap to vanish by mixing the singlet states 
with the spinful states. Alternatively, the experimental data
can be interpreted to imply that the KLHM
has no spin-gap in the thermodynamic limit.

We now turn our attention back to the pure KLHM. The log-log plot in Fig.~5
makes clear that around $T=J$, where HTE convergence fails, there is 
a crossover in the temperature dependence of the susceptibility. Below $T=J$, 
the susceptibility is remarkably well fit by a power law in temperature of the 
form $T^{-0.25}$. This non-trivial crossover behavior
between the high temperature Curie-Weiss regime and the very low temperature
RVB regime deserves further theoretical attention. Experimental
measurements of wavevector dependent susceptibilities, at temperatures
below $J$ can shed further light on this crossover.

\begin{figure}[!htb]
\begin{center}
  \includegraphics[scale=.55,angle=0]{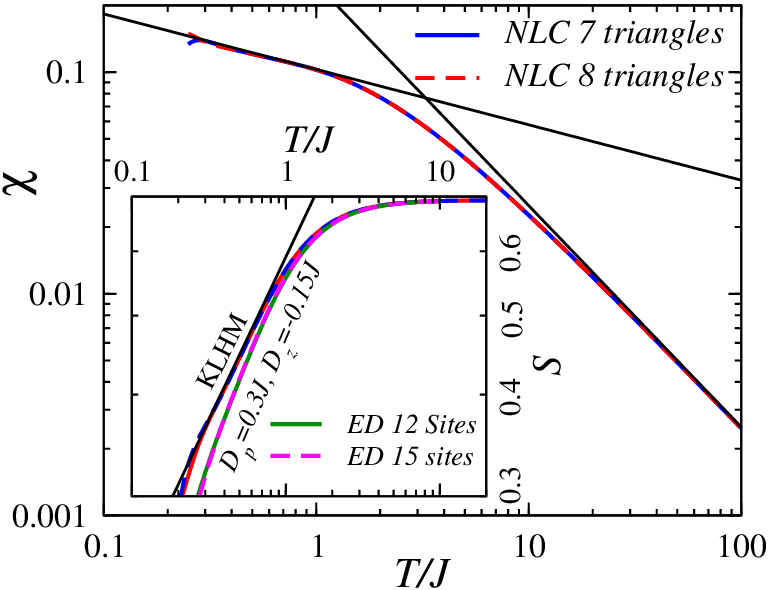}
\end{center}
\vspace{-0.6cm}
\caption{\label{Susceptibility_LogLog.eps} Log-log plot of the
susceptibility of the KLHM. In the inset we show log-log plots of the 
entropy of the pure KLHM and in the presence of DM terms.
Straight lines following the data are $\chi\sim T^{-0.25}$ (low $T$),
$\chi= 1/(4T)$ (high $T$), and $S\sim T^{0.4}$ (inset).
}
\end{figure}

A rough power law is also seen in the entropy function of the KLHM, 
which goes as $S\sim T^{0.4}$ (Inset in Fig.~5). This may help explain the 
failure of naive Pade extrapolations for the specific heat of the KLHM 
calculated by HTE \cite{elstner}. It was found that Pade approximations 
lead to specific heat curves, which when integrated give a finite ground 
state entropy. It was suggested that this meant the KLHM specific heat has 
two peaks as a function of temperature as discussed by Elser earlier 
\cite{elser}, and HTE simply fails to recover the second peak. However, 
there is no strong evidence to support the second peak. Our results suggest 
that instead of a second peak, there maybe a non-trivial power-law regime 
at low or intermediate temperatures which simple Pade extrapolations of 
HTE fail to capture. As shown in the inset in Fig.~5, the addition of DM 
terms to the KLHM produces an increase in the exponent of the entropy from 
$0.4$, bringing it in rather good
agreement with low temperature measurements \cite{expts2}. If 
the magnetic specific heat for the material \zncu\ can be extracted at 
higher $T$, the above issues can be further addressed.

In conclusion, in this paper we have studied the magnetic susceptibility for
the material \zncu\ by comparing it with theoretical calculations for KLHM.
We find that at high temperatures the susceptibility data is well described
by KLHM with $J=170K$. The sudden upturn in the experimental susceptibility 
below $T=75K$ is not consistent with KLHM but can be explained by 
Dzyaloshinsky-Moriya interactions. Further experiments, can verify or refute 
the existence of such terms by measuring the anisotropies in the susceptibility. 
We have also shown that the susceptibility of KLHM has an intermediate temperature 
regime where $\chi$ goes as $T^{-0.25}$. 
The combined experimental and numerical results 
raise a fundamental question about whether the KLHM has a spin-gap, or whether 
it has a gapless spin-liquid ground state \cite{ran-06,ryu}.

\begin{acknowledgments}

This work was supported by the US National Science Foundation, Grant 
No.\ DMR-0240918, DMR-0312261, and PHY-0301052. We are grateful to Oren Ofer, 
Amit Keren, Joel Helton, and Young Lee for providing us with the experimental 
susceptibility data, and to Takashi Imai for valuable discussions. 
Computational facilities have been provided by HPCC-USC center.

\end{acknowledgments}

{\it Note Added}.---Recent NMR experiments by Imai {\it et~al.} \cite{imai} 
find that parts of the NMR spectra show a temperature dependence like 
the bulk susceptibility that grows rapidly as $T$ is lowered while 
other parts flatten out and then decrease at still lower $T$. This is 
exactly what we find for $\chi_z$ and $\chi_p$ respectively. Thus a 
simple explanation for these experiments is that they arise from powders 
oriented perpendicular and parallel to the field.


\begin{thebibliography}{99}

\bibitem{expt-review} 
A. P. Ramirez {\it et~al.}, Nature {\bf 399}, 333 (1999);
S. T. Bramwell and M. J. P. Gingras, Science, {\bf 294}, 1495 (2001); 
S. H. Lee {\it et~al.}, Nature {\bf 418}, 856 (2002).

\bibitem{theory}
T. Senthil {\it et~al.}, Science {\bf 303}, 1490 (2004);
R. Moessner, S. L. Sondhi and E. Fradkin, 
Phys. Rev. B {\bf 65}, 024504 (2001).

\bibitem{lhuillier}
P. Lecheminant  {\it et~al.},
Phys. Rev. B {\bf 56}, 2521 (1997).

\bibitem{mila}
F. Mila,
Phys. Rev. Lett. {\bf 81}, 2356 (1998).

\bibitem{expts1} 
O. Ofer {\it et~al.}, 
cond-mat/0610540.

\bibitem{expts2} 
J. S. Helton {\it et~al.}, 
Phys. Rev. Lett. {\bf 98}, 107204 (2007). 

\bibitem{ran-06} 
Y. Ran {\it et~al.}, 
Phys. Rev. Lett. {\bf 98}, 117205 (2007).

\bibitem{nlc} 
M. Rigol, T. Bryant, and R. R. P. Singh,
Phys. Rev. Lett. {\bf 97}, 187202 (2006); 
to be published.

\bibitem{elstner} 
N. Elstner and A. P. Young,
Phys. Rev. B {\bf 50}, 6871 (1994).

\bibitem{zheng} 
W. Zheng {\it et~al.},
Phys. Rev. B {\bf 71}, 134422 (2005).

\bibitem{harris} 
T. Yildirim and A. B. Harris,
Phys. Rev. B {\bf 73}, 214446 (2006).

\bibitem{yamabe} 
Y. Yamabe {\it et~al.}, 
cond-mat/0607440.

\bibitem{elhajal02}
M. Elhajal, B. Canals, and C. Lacroix,
Phys. Rev. B {\bf 66}, 014422 (2002).

\bibitem{anisotropy} Assuming a comparable g-factor and DM anisotropy, 
the DM anisotropy needed to explain the experiments drops approaching 
Cs$_2$CuCl$_4$ (comparing $D$ and $J$ on the bonds on which 
DM terms are allowed by symmetry) \cite{coldea}.

\bibitem{coldea}
R. Coldea {\it et~al.}, 
Phys. Rev. Lett. 88, 137203 (2002).

\bibitem{grohol} 
D. Grohol {\it et~al.}, 
Nat. Mat. {\bf 4}, 323 (2005).

\bibitem{elser}
V. Elser,
Phys. Rev. Lett. {\bf 62}, 2405 (1989).

\bibitem{ryu} 
S. Ryu {\it et~al.}, 
cond-mat/0701020.

\bibitem{imai} 
T. Imai {\it et~al.}, 
cond-mat/0703141.

\end{thebibliography}
\end{document}